\documentstyle[12pt]{article}
\begin{document}
\begin{center}
{\bf \Large{A study of one dimensional correlated disordered systems using
invariant measure method}}\\
\vspace {.2in}

K. Kundu \footnote{email: kundu@iopb.ernet.in}, D. Giri
\footnote{email: giri@iopb.ernet.in} and K. Ray \footnote{email:
koushik@iopb.ernet.in} 

\vspace {.2in}

{\it Institute of Physics, Bhubaneswar 751005, India}\\
\vspace {.5in}
Abstract
\end{center}
The behavior of electronic states of one dimensional correlated
disordered systems which are modelled by a tight binding Hamiltonian
is studied analytically using the invariant measure method. The
approach of Bovier is generalized to include the possibility of
different site energies and nearest neighbor hopping integrals inside
the correlated sites or the cluster. The process is further
elaborated by applying to the symmetric random trimer model which
contains in it many hitherto known models of this category. An
alternative mathematical definition of the exceptional energy ($E_S$) from 
the invariant measure density along with physical arguments
substantiating it is presented. Furthermore, the procedure 
for obtaining exceptional energies is outlined and applied to the
symmetric random trimer model to derive conditions for obtaining
doubly degenerate exceptional energies.  The Lyapunov  
exponent ($\gamma (E)$) or the inverse localization length of states 
around the exceptional energy is found to vary as $\sim (E - 
E_S)^{2n}$ in the leading order. $n$ denotes the degeneracy of the
exceptional energy. The density of states at the exceptional
energies are calculated. We further propose that one dimensional
correlated disordered systems can be mapped to a Lloyd model in 
which the width 
of the distribution of site energies is determined by the reflection
coefficient of the cluster embedded in the lattice of the another
constituent. The importance of our results is discussed.

\newpage

\section{Introduction}
One of the well established results in condensed matter physics is
that all electronic eigenstates of a disordered one dimensional
system are exponentially localized irrespective of the strength of
the disorder. The early work of Anderson \cite{ander} on uncorrelated
site diagonal disorder in the tight binding model (TBM) and of Mott
and Twose \cite{mott} form the basis of this result. Of course, the
result of Anderson and Mott and Twose cannot be rigorously valid 
in one dimensional systems in which the disorder is correlated. For
example, in the context of a TBM it has been shown that correlated
off-diagonal disorder \cite{econ} cannot localize the state at the
band center. Another example in this category is the model proposed
by Dunlap, Kundu and Phillips (DKP) \cite{dkp}. The most well known
example in the context of a TBH, however, is the random dimer model
(RDM) \cite{rdm}. This is basically the offspring of the original 
DKP model. The generalization of the RDM requires the extension of
the correlation beyond the nearest neighbor site and the
introduction of different nearest neighbor hopping elements among the
correlated sites or simply the cluster. Two good but simpler examples
in this generalized category are the repulsive binary alloy (RBA)
\cite{rba} and the symmetric random trimer model (SRTM) \cite{rtm}. 
The SRTM which is the further generalization of the RBA, yields two
exceptional energies. The exceptional energy is the energy at which
the reflection coefficient of the cluster embedded in the lattice of
the other constituent vanishes. We further note that the number of
exceptional energy for a trimer embedded in a one dimensional lattice
of another element cannot exceed two. Another interesting
as well as important feature of the SRTM is that positions of these 
exceptional 
energies can be tuned by changing either the hopping element or the 
site energies of the cluster and these two energies 
merge as a limiting case. 
This has actually been shown by Giri, Datta and Kundu (GDK) 
\cite{rtm} by conventional analysis of the reflection coefficient and
by appropriate numerical simulations. Salient features of the SRTM 
can be verified by fabricating appropriate layered heterojunction 
and also by coupling quantum dots \cite{dot}. Another potential area of
application of this model is the field of organic conducting polymers
\cite{poly}.

The most commonly used method for studying the electronic properties
of one dimensional correlated systems is to analyze the reflection 
coefficient of the cluster in the neighborhood of exceptional
energies. Since the system behaves like a weakly disordered system in
the neighborhood of these energies, a very good estimation, albeit
not rigorous, of the Lyapunov exponent (inverse localization length)
in these neighborhoods can be obtained from the reflection
coefficient.  On the other hand an estimation of the density of
states (DOS) at an exceptional energy can in principle be obtained 
from the phase of the transmission coefficient of a single cluster 
through the Thoules' formula \cite{thou}. To the best of our
knowledge no such effort has been made in this direction. 
Hence, the DOS and the mean square displacement of the 
particle are calculated numerically \cite{rtm1} to establish the 
presence of nonscattered states around these energies. 

Another way of looking at this state of affairs is the calculation of
the Lyapunov exponent and the integrated density of states (IDOS) of
the system from its invariant measure. In an attempt toward
understanding the behavior of the IDOS of the one dimensional
Anderson model in the weak disorder limit, Bovier and Klein
\cite{bovk,bov0} developed a scheme for a perturbation expansion of the 
invariant measure of the model. From the modified perturbative
expansion of the invariant measure, Bovier and Klein showed that at
all energies $E_0 = 2 \cos \alpha \pi$ with $\alpha$ rational, the
IDOS of the Anderson model in the weak disordered limit has
singularities. This was the extension of the results obtained
previously by Kappus and Wegner \cite{kap} and Derrida and Gardner
\cite{derr}. Bovier and Klein \cite{bovk} further showed that for
irrational $\alpha$, their technique gives a unique invariant measure
with finite coefficients to all orders of perturbation. This modified
expansion has also proved to be a true asymptotic expansion of the
invariant measure \cite{camp}. This scheme was later applied by
Bovier \cite{bov} 
to develop the perturbation series expansion of the invariant measure
around the exceptional energies of the RDM. This enabled him to show
that the Lyapunov exponent vanishes as $\varepsilon^2$ in energy
($\varepsilon$) in the neighborhood of the exceptional energies.
Furthermore, the IDOS is found to vary as $\varepsilon$ within this
energy width. This is basically the first rigorous calculation on the
RDM confirming the results of Ref. \cite{rdm}. 

The fundamental characteristic of the cluster correlated disordered
systems is the presence of exceptional energies where disorder systems
purportedly behave like perfect systems. However, for these systems
to play an important role in transport properties of the materials,
there must be a finite DOS at these energies. Hence, to fully
characterize these systems, we need rigorous analytical calculations 
of the Lyapunov exponent around these energies and DOS at these
energies. To the best of our knowledge, for these systems, the
invariant measure technique is the only technique that can yield 
the DOS analytically without invoking any approximation. This is 
primarily the motive to apply this technique to the SRTM which 
encompasses many hitherto known examples. For this purpose, we 
generalize the approach of Bovier \cite{bov} to 
include the possibility of different site energies and nearest 
neighbor hopping 
elements in the cluster. We also give here an alternative mathematical 
definition of the exceptional energy from the invariant measure and
derive from it in the case 
of the SRTM, an algebraic equation in energy involving relevant 
parameters of the system. We further show that the equation gives 
the correct prediction of the possibility of tuning of exceptional 
energies yielding in the limit a degenerate exceptional energy 
as noted by GDK \cite{rtm}. Finally we propose a mapping of these 
systems to an effective Lloyd model. Such a mapping will
be useful in the study of transport properties of these systems. 

The organization of the paper is as follows. In the following section
we generalize the approach of Bovier. We then develop equations for
the invariant measure density, the Lyapunov exponent and the IDOS. In
section 4 we discuss the method for obtaining exceptional energies .
Section 5 is devoted to the perturbative calculation of the invariant
measure density for the SRTM. In section 6 and section 7 we calculate
the leading order behavior of the Lyapunov exponent and the DOS at
exceptional energies respectively. In section 8 we deal with the
mapping of the aspect. We conclude the paper by highlighting the
major contributions of the paper.

\section{Formalism}
We study the Hamiltonian
\begin{equation}
H = \sum_{n} a_n^{\dagger} a_n + \sum_n V_{n+1, n} (
a_{n+1}^{\dagger} a_n + a_n^{\dagger} a_{n+1})
\end{equation}
on $l^2 (Z)$ where $a_n (a_n^{\dagger})$ destroys (creates) a
particle at the $n$th site. $V_{n, n+1}$ is the tunneling matrix
connecting the $n$th site to the $(n+1)$th site. $\{ V_{n+1, n} \}$ are
taken to be real and positive, although this constraint is not
necessary for the formalism.  

The eigenvalue equation \cite{ecbook} associated with $H$ is 
\begin{equation}
\epsilon_n C_n + V_{n+1, n} C_{n+1} + V_{n,n-1} C_{n-1} = E C_n
\end{equation}
We introduce $z_n = \frac{V_{n,n-1} C_n}{C_{n-1}} \in {\dot {\bf R}}$
with $\dot {\bf R}$
denoting the compactified real line ${\bf R} \cup \{\infty\}$. The
recursion relation for $z_n$ is then 
\begin{equation}
z_{n+1} = E - \epsilon_n - \frac{V_{n,n-1}^{2}}{z_n} \equiv \xi_{E,
\epsilon_n V_{n,n-1}} (z_n)
\end{equation}
We further note in this connection that the eigenvalue equation 
for a one dimensional array of masses $\{m_{i}\}$ coupled to nearest
neighbors by identical harmonic springs is 
\begin{equation}\label{mass}
(2 - m_i \Omega^2) u_i = u_{i-1} + u_{i+1}
\end{equation}
where $u_i$ is the displacement of the $i$th mass, $m_i$ in the vibration 
with frequency $\Omega$. All spring constants are taken to be unity
without any loss of generality. Now introducing a variable, $z_n =
\frac{u_n}{(m_{n-1} u_{n-1})} \in {\dot {\bf R}}$, we obtain from
(\ref{mass}) 
\begin{equation}
z_{n+1} = -\Omega^2 + \epsilon_n - \frac{V_{n,n-1}^2}{z_n}
\end{equation}
where $\epsilon_n = \frac{2}{m_n}$ and $V_{n,n-1} = (m_n
m_{n-1})^{-1/2}$. 
Hence, the behavior of one dimensional array of masses coupled by
harmonic springs is mathematically equivalent to the one dimensional
quantum motion of a particle in a TBM \cite{mat,phd}.  The Lyapunov
exponent, $\gamma 
(E)$ and the IDOS, $N(E)$ are related to the large $n$ behavior of
$z_n$. 
If we define 
\begin{equation}
\tilde \gamma (E) = \lim_{N\rightarrow \infty} \frac{1}{N}
\sum_{n=1}^{N} \ln \frac{z_n}{V_{n,n-1}}
\end{equation}
then 
\begin{equation}
\gamma (E) = {\bf \rm Re} \; \tilde \gamma (E)
\end{equation}
and 
\begin{equation}\label{nfc}
N (E) = \frac{1}{\pi} {\bf \rm Im} \; \tilde \gamma (E)
\end{equation}

To understand the origin of equation (\ref{nfc}) we consider a chain of
$N$ sites with fixed boundary conditions at both ends. In other words
we set $C_{-1} = C_N = 0$. To keep the argument simple, we assume that
$\{V_{n, n-1} \}$, $n \in Z$ are real. From equation (3) we get $z_1 = E -
\epsilon_1$, a real quantity. So, none of the $\{z_n \}$ can be truly
complex. However, $\{ \frac{Z_n}{V_{n,n-1}} \}$, $n \in Z$ can be
negative real numbers which can be thought of as complex numbers with
the minimum phase, $\pi$. So, the right hand side of equation (\ref{nfc})
picks up a contribution whenever $\frac{z_n}{V_{n,n-1}} =
\frac{C_n}{C_{n-1}}$ is negative. 

Consider now the case when $C_n = 0$ for $n \leq N$. From equation (3) we 
obtain 
\begin{equation}\label{new}
z_{n-1} = \frac{V_{n-1,n-2}^2}{E-\epsilon_{n-1}} = E - \epsilon_{n-2}
- \frac{V_{n-2,n-3}^2}{z_{n-2}}
\end{equation}
Values of $E$ for which this equation (\ref{new}) is satisfied are the
eigenvalues of the system. When $n = N$, we obtain eigenvalues 
of the system under study. Let us assume that $E_l$ is the $l$-th
eigenvalue of the system in ascending order. So, the eigen vector 
belonging to $E_l$ has $(l-1)$ nodes. This in turn implies that 
$\frac{z_n}{V_{n,n-1}}$ will also pick up $(l-1)$ negative numbers
giving ${\rm Im} \frac{1}{N} \sum_{n=1}^{N} \frac{z_n
(E_l)}{V_{n,n-1}} = \frac{l-1}{N}$. On the other hand, the number of
sates upto $E = E_l$, {\it i.e} IDOS ($E_l$) is $l$. So, in the limit
$N \rightarrow \infty$, equation (\ref{nfc}) yields the IDOS ($E_l$). 
To understand further the behavior of equation (\ref{nfc}) for $E_l < E <
E_{l+1}$, we note that $E_l (m) > E_l (N)$ if $m < N$. For $E$ in
this limit equation (\ref{new}) will be satisfied for some $m$ such that 
$l \leq m < N$ and the $l$-th eigen value of the reduced system will
be obtained. Hence, the number of modes and consequently the number
of negative values of ($\frac{Z_{n}}{V_{n,n-1}}$) will be preserved. 
We further note that when $E = E_l + \epsilon$ and $\epsilon
\rightarrow 0$, $m$ will be close to $N$. When $E = E_{l+1} -
\epsilon$, $m$ will be close to $l$ and it will swing back to $N$ for
$E = E_{l+1}$. For further discussion on this see ref.\cite{dean}.

The disorder in the Anderson model can arise from the disorder in 
diagonal elements ($\epsilon_n$) of $H$, from the disorder in
off-diagonal elements, $V_{n,n+1}$, $n \in Z$ of $H$ or from both. In all
these cases equation(3) defines a Markov chain in which states are
characterized by the random variable $z_n$, $n \in Z$. When
$V_{n,n+1} = V$, $n \in Z$ and $\{ \epsilon_n \}$ are  i.i.d
random variables, this Markov chain consists of persistent non-null
states \cite{fell}. In other words, the chain is ergodic.
Furstenberg's theorem \cite{fur} then asserts that $\epsilon_n \neq
\tilde \epsilon$, $\tilde \epsilon$ = a constant, 
$n \in Z$, there is a unique invariant measure $d\nu_E (z)$ on $\dot
{\bf R}$. This measures satisfies
\begin{equation}
\int_{\dot {\bf R}} d \nu_E (z) f(z) = {\bf \rm E }\int_{\dot {\bf
R}} d\nu_E (z) 
f(E -\epsilon -\frac{1}{z})
\end{equation}
for all bounded measurable functions, $f$. Here, {\bf E} denotes the
expectation with respect to the probability distribution of
$\epsilon$. Furthermore, this measure is actually continuous and
hence has support on {\bf R}. So, the measure $ d\nu_E (z)$ has a
density, {\it i.e.}, $d\nu_E (z) = \phi_E (z) dz$, where $\phi_E (z)$
defines the density at $z \in {\bf R}$. Similarly, if $\epsilon_n =
\tilde \epsilon$, $n \in Z$ and $\{ V_{n,n+1} \}$ are  i.i.d random
variables, the resulting Markov chain is also ergodic except at 
$E = \tilde \epsilon$. Hence, in this case also a unique invariant
measure exists on $\dot {\bf R}$ \cite{bov0}. From these established 
results we conclude that 
when both $\{ \epsilon_n \}$ and $\{ V_{n,n+1} \}$ are  i.i.d
random variables, the resulting Markov chain also consist of
persistent non-null states. So, again a unique invariant measure will
exist on $\dot {\bf R}$.

To apply these results to correlated disordered systems some
modifications are needed. In order to introduce the required
modifications, we first describe the model. The model considered here
is a random binary mixture of two types of clusters. Each cluster
contains $q \geq 2$ elements. In the host cluster all elements are
assumed to be same while in the guest cluster at least one element, 
if not all, should be distinct from the host element. All site 
energies and 
nearest neighbor hopping integrals in the host cluster are set to 
zero and unity 
respectively without any loss of generality. In the guest cluster
site energies and nearest neighbor hopping integrals are allowed to be
different. We further assume that the hopping integrals between the
end sites of any two clusters are unity. 
So, diagonal and off-diagonal elements in our model are not totally
random. Instead each category is required to satisfy $q$ constraint
relations :
\begin{equation}\label{10}
\tilde \epsilon_{qm + l - 1} = [\epsilon_{l-1} \; p_{qm} +
\epsilon_{q-l} \; (1-p_{qm})] e_{qm}
\end{equation}
and
\begin{equation}\label{11}
\widetilde{V}_{qm + l, qm + l-1} = 1 - [1- (V_{l,l-1} \; p_{qm} +
V_{q-1-l,q-l} \; (1-p_{qm}))] (1-\delta_{q,l}) e_{qm}
\end{equation}
where $1 \leq l \leq q$ and $qm \in Z$. The randomness in the model
is, therefore, introduced through two Rademacher variables, $e_{qm},
p_{qm} \in \{ 0, 1\}$. These variables by construction are indeed
i.i.d random variables. Rademacher variables, $\{ p_{qm} \}$ are 
introduced 
to take into account the asymmetry of the guest cluster. Since a
cluster can take only two possible orientations in the lattice, two
possible values of $\{ p_{qm} \}$, namely zero and unity, occur with
probability $\frac{1}{2}$. 

Since diagonal elements of $H$ here determine the strength of
hopping to and from the sites, the off-diagonal elements are not
truly random. In essence this model is similar to the Anderson model
of uncorrelated site disorder. 
However, due to constraints on diagonal and off-diagonal elements
of $H$, $\{ Z_n \}$ as such do not form a Markov chain. To form the 
required 
Markov chain we need to consider the clusters as unit cells. 
In other words, we need to define a new random
variable $\{ X_n \}$ such that
\begin{eqnarray}
x_{n+1} & = & z_{q(m+1)} \nonumber \\
& = & \prod_{l=1}^{q} \xi_{E, \epsilon_{q (m+1) -l}, \tilde V_{q
(m+1) - l, q (m+1) - (l+1)}} (x_n)  \label{xn} \\
& \equiv & {\bf P}_{E,q} (x_n) \nonumber
\end{eqnarray}
We note that in the product of the operators in equation (\ref{xn}), the
operator with the lower value of $l$ comes to the left. The Markov chain
defined by the random variable, $\{ X_n \}$ is ergodic and
according to Frustenberg's theorem, a unique invariant measure
$d\nu_E (x)$ for this process will exist on $\dot {\bf R}$.
Furthermore, this measure will have density, {\it i.e.} $d \nu_E (x)
= \phi_E (x) dx$, when $\phi_E (x) \in L_{+}^{1} (\dot {\bf R},\; dx)$. 
In other words $\phi_E (x)$ belongs to the class of
nonnegative Lebesgue integrable functions on the compactified real
line.  

\section{Equations for the invariant measure $\phi_E (x)$, the
Lyapunov exponent and the Integrated density of states}
Before defining the equation for determining $\phi_E (x)$, we define 
an operator, ${}^{V_{0}} T_{E}$ such that
\begin{equation}
({}^{V_0} T_{E} f) (x) = \frac{V_0^2}{(E-x)^2} f (\frac{V_0^2}{E-x})
\end{equation}
This is the logical extension of the definition given by Bovier and
Klein \cite{bovk}. This generalization is useful for the more general
problem as it can be seen in the forth coming discussion. 
Some important properties of ${}^{V_0} T_E$ which will be used here
are 

\vspace {.25in}

(i) ${}^{V_0} T_{E-v} {}^{V_0} T_{E-u} = T_{E-v}
T_{\frac{E-u}{V_0^2}}$ \\ 

(ii) ${}^{V_0} T_E^{-1} \frac{d}{dx} \; {}^{V_0} T_E = \frac{1}{V_0^2}
\frac{d}{dx} x^2$, including $E = 0$. \\

(iii) ${}^{V_0} T_0 \; {}^{V_0} T_0 = {\rm I}$ \\

(iv) ${}^{V_0} T_E = e^{-E \frac{d}{dx}} \; \; {}^{V_0} T_0$, \\

(v) $ ({}^{V_0} T_E^{-1} f ) (x) = \frac{V_0^2}{x^2} f (E -
\frac{V_0^2}{x})$, \\

If $d\nu_E (x)$ defines the invariant measure with respect to the
process described by $\{ X_n \}$, introducing equation (13) in
equation (10) we obtain
\begin{equation} \label{eqn}
\int_{\dot {\bf R}} d\nu_E (x) f(x) = {\bf \rm E} \int_{\dot {\bf R}}
d\nu_E (x) f({\bf P}_{E,q} (x) )
\end{equation}
for all bounded measurable functions. As usual {\bf E} denotes
expectation with respect to the probability distribution of $\{ e_{q
m} \}$ and $\{ p_{qm} \}$. In our model calculations we take 
\begin{equation}
P(e) = \frac{1}{2} \delta (e) + \frac{1}{2} \delta (e-1)
\end{equation}
Since $d\nu_E (x) = \phi_E (x) dx$, introducing this relation in equation
(\ref{eqn}) and making appropriate change of variables we obtain 
\begin{equation}\label{eqn1}
\int_{\dot {\bf R}} \phi_E (x) dx = {\bf \rm E} \int_{\dot {\bf R}} dx
\prod_{l=1}^{q} \; {}^{\widetilde{V} (q,l)}T_{E-\epsilon (q,l)} \;
\phi_E (x) 
\end{equation}
where $\epsilon (q,l)$ and $\tilde{V} (q,l)$ are defined respectively
by equation(\ref{10}) and (\ref{11}) by setting $p_{qm} = p$ and $e_{qm} =
e$. Again in the product of these operators, the operator with lower
value of $l$ comes to the left. Since equation (\ref{eqn1}) should hold
good for any arbitrary bounded measurable function, $f$, we obtain 
after averaging over $e$ and $p$
\begin{equation}
\phi_E (x) = [\frac{1}{2} T_{E}^{q} + \frac{1}{4} \prod_{l=1}^{q} \; 
{}^{V_{q-l,q-(l+1)}} T_{E-\epsilon_{q-l}} + \frac{1}{4}
\prod_{l=1}^{q} \; {}^{V_{l-1,l}} T_{E-\epsilon_{l-1}}] \phi_E (x) 
\end{equation}
As an example we consider the SRTM. For this model $q =3$,
$\epsilon_2 = \epsilon_0 = v$ and $ \epsilon_1 = u$. The hopping
between nearest-neighbors in the guest cluster is $V_0$. Since the 
cluster has 
a inversion symmetry, two products are identical. So, we have 
\begin{eqnarray}
\phi_E (x) & = & \frac{1}{2} [ T_E^3 + {}^{V_0}T_{E-v} \; {}^{V_0}T_{E-u}
\; T_{E-v} ] \phi_E (x) \nonumber \\
& = & \frac{1}{2} [ T_E^3 + T_{E-v} T_{(E-u)/V_0^2} T_{E-v} ] \phi_E
(x) \label{20}
\end{eqnarray}

Before computing the complex Lyapunov exponent, $\tilde \gamma (E)$,
we note that for disordered systems including systems under study, due
to the subadditive ergodic theorem \cite{bov0,rue} the limit in
equation(6) 
exists and is independent of the realization of the disorder, for all
most all realization. In other words, $\tilde \gamma (E)$ is self
averaging,
\begin{equation}
{\bf \rm E} \tilde \gamma (E) = \tilde \gamma (E)
\end{equation}
where {\bf E} denotes ensemble average. For effective calculation of 
$\tilde \gamma (E)$, we first write 
\begin{equation}
\tilde \gamma (E) = \lim_{N\rightarrow \infty} \tilde \gamma_N (E)
\end{equation}
where $N = q M$ is the total number of sites in the chain. Since 
\begin{eqnarray}
z_{qm +l} & = & \prod_{k=1}^{l} \xi_{E,\epsilon (q, l-k), \tilde V (q,
l-k)} (z_{qm}) \nonumber \\
& = & P_{E,l} (x_n)
\end{eqnarray}
[see equation (13)], from equation (6) we obtain
\begin{equation} \label{qq}
\tilde \gamma_N (E) = \frac{1}{q M} \sum_{n=0}^{M} \ln x_n +
\frac{1}{q M} \sum_{n=0}^{M} \sum_{l=1}^{q-1} \ln P_{E,l} (x_n) -
\frac{1}{q M} \sum_{n=0}^{M} \ln \prod_{l=0}^{q-1} \tilde V_{q n + l,
q n + l - 1}
\end{equation}
The last term in equation (\ref{qq}) when averaged over all possible
realizations of the sample (over $e$ and $p$) yields 
\begin{equation}
< \frac{1}{q M} \sum_{n=0}^{M} \ln \prod_{l=0}^{q-1} \tilde V_{q n + l,
q n + l - 1} >_{e, p} = \frac{1}{2 q} \ln \prod_{l=1}^{q-1} V_{l-1,l}
\end{equation}
In the limit $N \rightarrow \infty$ we apply
the subadditive ergodic theorem to equation (\ref{qq}). This in turn
yields 
\begin{eqnarray}
\tilde \gamma (E) & = & \frac{1}{q} \int_{\bf \rm \dot R} d\nu_E (x)
\ln x \nonumber \\ 
 &  & + \frac{1}{q} \sum_{l=1}^{q-1} {\bf \rm E} \int_{\bf \rm \dot R}
d\nu_E (x) \ln P_{E,l} (x) - \frac{1}{2 q} \ln \prod_{l=1}^{q-1} V_{l-1,l}
\label{28}
\end{eqnarray}
where {\bf E} denotes expectation over $\{ e \}$ and $\{ p\}$.
Furthermore, introducing $d\nu_E (x) = \phi_E (x) dx$ and 
making appropriate change of variables in the second integral of equation
(\ref{28}), we obtain 
\begin{eqnarray}
\tilde \gamma (E) & = & \frac{1}{q} \int_{\bf \rm \dot R} dx \phi_E
(x) \ln x \nonumber \\ 
&  & + \frac{1}{q} \sum_{l=1}^{q-1} {\bf \rm E} \int_{\bf \rm \dot R}
dx \ln x \prod_{m=0}^{l} \; {}^{\tilde V (q,l-m)} T _{E-\epsilon (q,
l-m)} \phi_E (x)  - \frac{1}{2 q} \ln \prod_{l=1}^{q-1} V_{l-1,l}
\end{eqnarray}
We again point out that in the product of the operators, the operator
with the lowest value of $m$ comes to the left. Furthermore, the
product terminates at $m =l$. Equation (26) 
when averaged over $e$ and $p$ yields 
\begin{eqnarray}
\tilde \gamma (E) & = & \frac{1}{q} \int_{\bf \rm \dot R} dx \phi_E
(x) \ln x +  \frac{1}{2q} \sum_{l=1}^{q-1} \int_{\bf \rm \dot R}
\sum_{l=1}^{q-1} dx \ln x \; T_{E}^{l} \phi_E (x) \nonumber \\
&  & + \frac{1}{4 q} \sum_{l=1}^{q-1} \int_{\bf \rm \dot R} dx \ln x 
\prod_{m=1}^{l} \; {}^{V_{l-m, l-m-1}} T _{E-\epsilon_{l-m}} \phi_E
(x)\nonumber \\
&  & + \frac{1}{4 q} \sum_{l=1}^{q-1} \int_{\bf \rm \dot R} dx \ln x 
\prod_{m=1}^{l} \; {}^{V_{q-1-l+m, q-l+m}} T _{E-\epsilon_{q-l+m-1}}
\phi_E (x) \nonumber \\ 
&  & - \frac{1}{2 q} \ln \prod_{l=1}^{q-1} V_{l-1, l} \label{30}
\end{eqnarray}
Before applying equation (\ref{30}) to the SRTM we first observe that 
\begin{equation}
\int_{\bf \rm \dot R} dx \ln x \; {}^{V_0} T_{E-u} \; T_{E-v} \phi_E
(x) - \ln V_0^{2} = \int_{\bf \rm \dot R} dx \ln x  \; \;
T_{\frac{(E-u)}{V_0^2}} \; T_{E-v} \phi_E (x)
\end{equation}
Decomposing the complex Lyapunov exponent $\tilde \gamma (E)$ to real
and imaginary parts, we obtain for the IDOS, $N(E)$ of the SRTM
\begin{eqnarray}
N (E) & = & 1 - \frac{1}{\pi} {\rm Im} \tilde \gamma (E) \nonumber \\
& = & 1 - \frac{1}{6} \int_{-\infty}^{0-} dx  [2 + T_E + T_E^2 + T_{E-v}
+ T_{\frac{(E-u)}{V_0^2}} \; T_{E-v} ] \phi_E (x)
\end{eqnarray}
Furthermore, for the Lyapunov exponent $\gamma (E)$ we get
\begin{eqnarray}
\gamma (E)  & = & {\rm Re} \tilde \gamma (E) \nonumber \\
& = & \frac{1}{6} \int_{-\infty}^{\infty} dx \ln y(x) [
T_{E-v} \; T_{\frac{(E-u)}{V_0^2}} \; T_{E-v} - T_E ] \phi_E (x)
\nonumber \\
&  & + \frac{1}{6} \int_{-\infty}^{\infty} dx \ln \mid x \mid [ T_{E-v}
+ T_{\frac{(E-u)}{V_0^2}} \; T_{E-v} ] \phi_E (x) \label{32}
\end{eqnarray}
To obtain the first integral of equation (\ref{32}) we define 
$y (x) = \sqrt{x^2 - E x + 1}$ so that $\mid x
\mid = y(x) / y (E - \frac{1}{x})$. 
This in turn yields 
\begin{equation}
\ln \mid x \mid = \ln y(x) - \ln y (E - \frac{1}{x}) \label{33}
\end{equation}
and the integral decomposes into two integrals. In the second
integral we replace $x$ by $1/(E-x)$ so that $y (E-\frac{1}{x})
\rightarrow y (x)$. Then we combine these integrals and introduce the
governing equation of $\phi_E (x)$ [equation (\ref{20})] \cite{bov} for
the final result. We 
further note that the effect of the guest cluster on $\gamma (E)$
appears in the second integral of equation (\ref{32}). So, the
transformation required for this integral will depend on the
structure of the guest cluster. For example, consider the example of
the RBA. Here, $v = 0$ and $u = (1-V_0^2) \omega$. For this case we
simply introduce equation (\ref{33}) in equation (\ref{32}) and after 
some algebra we obtain 
\begin{equation}
\gamma (E) = \frac{1}{6} \int_{-\infty}^{\infty} dx \ln y (x) (
T_{\frac{(E-u)}{V_0^2}}  - T_E ) \; T_E \phi_E (x) \label{34}
\end{equation}
equation (\ref{34}) immediately yields $\gamma (\omega) = 0$ 
confirming the 
result of ref. \cite{rba}. As a second example, we consider $u = (1 +
V_0^2) v$ in the SRTM. In this case the system has an exceptational
energy at $E = v$. To obtain the behavior of $\gamma (E)$ in the
neighborhood of $v$ we transform equation (\ref{32}) to 
\begin{eqnarray}
\gamma (E)  & = & \frac{1}{6} \int_{-\infty}^{\infty} dx \ln y(x) [
T_{E-v} \; T_{\frac{(E-u)}{V_0^2}} \; T_{E-v} - T_E ] \phi_E (x)
\nonumber \\
&  & - \frac{1}{6} \int_{-\infty}^{\infty} dx \ln y (x) [ 1- T_{E}] 
\; T_0 \; [T_{E-v} + T_{\frac{(E-u)}{V_0^2}} \; T_{E-v} ] \phi_E (x)
\label{35} 
\end{eqnarray} 
Since $T_0^2 = {\rm I}$ and $T_0 \; T_{-v} \; T_0 = T_v^{-1}$, from
equation (\ref{35}) we obtain $\gamma (v) = 0$ which is in
agreement with the
result in ref \cite{rtm}. When $V_0 = 1$, this system also has
another exceptional energy at $E = 2 v$. The calculation of $\gamma
(E)$ for this case is facilitated by converting equation (\ref{32}) to 
\begin{eqnarray}
\gamma (E)  & = & \frac{1}{6} \int_{-\infty}^{\infty} dx \ln y(x) [
T_{E-v} \; T_{E- 2v} \; T_{E-v} - T_E ] \phi_E (x)
\nonumber \\
&  & + \frac{1}{6} \int_{-\infty}^{\infty} dx \ln \mid x \mid \;
[T_{E-v} - T_0 \; T_{E-2v} \; T_{E-v}] \phi_E (x) \label{36}
\end{eqnarray}
We note that when $E = 2 v$, $T_{E-v} \; T_{E- 2v} \; T_{E-v} = T_v
\; T_0 \; T_v = T_{2v}$ and $T_0 \; T_{E- 2v} \; T_{E-v} = T_v$. So, 
$\gamma (2 v) = 0$. This result has also been obtained in ref.
\cite{rtm}. There is another important motivation behind bringing
$\ln y (x)$ in the integrals. When $E = 2 \cos\alpha \pi$, these
integrals can be easily carried out by mapping the compactified real
line on the circumference of a circle of circumference, $\pi (S_1)$
through the transformation, 
\begin{equation} \label{x}
x = \frac{\sin(\theta + \alpha \pi)}{\sin \theta}, 
\end{equation}
when $x \in {\bf \dot R}$ and $\theta \in S_1$.
As it will be shown later, this transformation will also simplify the
perturbative calculation of $\phi_E (x)$.

\section{Exceptional energy}
For the purpose of clarity we discuss this aspect in reference to the
SRTM. We already obtained the governing equation of $\phi_E (x)$ for
this model [equation (\ref{20})]. We note that for the RBA, at $E =
\omega$ equation (\ref{20}) reduces to 
\begin{equation}
\phi_{\omega} (x) = T_{\omega}^3 \; \phi_{\omega} (x) \label{p1}
\end{equation}
For the SRTM with $u = (1+V_0^2) v$ we obtain at $E = v$ 
\begin{equation}
\phi_{v} (x) = \frac{1}{2} [ T_v^3 + T_v^{-1} ] \phi_v (x) \label{p2}
\end{equation}
and for $V_0 = 1$ at $E = 2 v$, we get
\begin{eqnarray}
\phi_{2 v} (x) & = & \frac{1}{2} [T_{2 v}^3 + T_v \; T_0 \; T_v ]
\phi_{2 v} (x) \nonumber \\
& = & \frac{1}{2} [T_{2 v}^2 + {\rm I}] \; T_{2v} \; \phi_{2 v} (x)
\label{p3} 
\end{eqnarray}
We note now that $T_E \; f = f$ always has the normalized solution 
\begin{equation} 
f_E (x) = \frac{1}{\pi} \frac{\sqrt{1 - E^2 /4}}{x^2 - E x + 1}
\label{f0} 
\end{equation}
and $f (x) \in L_{+}^{1} (\dot R, dx)$ iff $\mid \frac{E}{2} \mid
\leq 1$. For $E = 2 \cos \alpha \pi$, if $\alpha$ is irrational,
this solution is unique \cite{bovk,bov}. 

It is important to note that the solution $\phi_E (x)$ of equation
(\ref{p1}), (\ref{p2}) and (\ref{p3}) is $f_{E_S} (x)$ when $E_S =
\omega$, $v$ or $2 v$. Keeping this in mind, we define the
exceptional energy as the energy at which $\phi_{E_S} = f_{E_S} (x)$.
This definition has also a physical origin. Consider two periodic
structures, one from the host cluster and the other from the guest
cluster. The periodic system from the guest cluster will in principle
form $q$ bands. Intersection of the DOS of any one of these bands
with the DOS of the periodic system from the host cluster determines 
the exceptional energy ($E_S$). Since the invariant measure density 
of the host system is given by equation (\ref{f0}), the definition ensues. 
Then, to derive the equation for exceptional energies of the SRTM we
need to solve 
\begin{equation}
f_E (x) = T_{E-v} \; T_{\frac{(E-u)}{V_0^2}} \; T_{E-v} f_{E} (x)
\label{f1} 
\end{equation}
It is relevant at this point to note that if 
\begin{equation}
g (x) = \frac{1}{\pi} \; \frac{a}{(x-b)^2 + a^2} \label{l1}
\end{equation}
then $T_E \;  g(x)$ is also another Lorentzian distribution centered at
$b_1$ with a half-width $a_1$, where
\begin{equation}
b_1 + i a_1 = E - \frac{1}{b + i a} \label{l2}
\end{equation}
Hence, from equation (\ref{f1}) we obtain
\begin{equation}
S = E - v - \frac{1}{\frac{E-u}{V_0^2} - \frac{1}{E - v
-\frac{1}{S}}} \label{s}
\end{equation}
where $S = \frac{E}{2} + i \sqrt{1 - E^2 /4}$.  equation (\ref{s}) in turn
yields a quadratic equation in $S$. The required equation is obtained
from either ${\rm Re} S = \frac{E}{2}$ or from ${\rm Im} S = \sqrt{1
- E^2 /4}$. Both will yield identical equations. This procedure for
the SRTM yields
\begin{equation}
v (E_S - v)^2 - [v(u-v) + (1-V_0^2) ] (E_S -v) + [u - (1+V_0^2) v] =
0 \label{gg}
\end{equation}
This equation has already been obtained in ref \cite{rtm} by
considering the reflection coefficient of a single guest cluster 
embedded in the host lattice. Full analysis of this equation can also
be found there. We note that for $v =0$ and $u = (1
-V_0^2) \omega$, from equation (\ref{gg}) we obtain $E_S = \omega$. On the
other hand if we take $u = (1 + V_0^2) v$, from equation (\ref{gg}) we
obtain $E_{s_1} = v$ and $E_{s_2} = (1 + V_0^2) v + (1-V_0^2) /v$. 
From these we can immediately see that two exceptional energies in
this case will
merge at $v$ if $V_0^2 = \frac{1}{1-v^2}$. For $V_0 = 1$,
similar situation can be obtained by setting $u = [v + \frac{2 [ 1 -
\sqrt{1-v^2} ]}{v}]$ and $\mid v \mid \leq 1$. For this case, two 
exceptional 
energies will coincide at $E_S = (u + v)/2$. 

We further note that the
procedure outlined here can be applied to complicated one dimensional
chains like polyaniline, polythiophene etc. We need to apply first
the real space renormalization procedure to these systems to obtain
effective one dimensional chains \cite{poly}. Then we can apply this 
procedure for finding exceptional energies in these systems. 

When the guest
cluster is asymmetric $f_E (x)$ has to satisfy two equations for two
different orientations of the cluster. This in turn will yields two
equations for $S$ and these need to be identical. This develops
further constraints in the parameter space. Because of this extra
constraint on the equation for $S$, the probability of obtaining
exceptional energies with asymmetric guest cluster, particularly if
the cluster size is relatively large, is negligible. One example of
an asymmetric guest cluster having an exceptional energy has been
worked out in ref. \cite{rtm}. This shows the importance of symmetry
in the guest cluster for obtaining exceptional energies. This aspect
has also been discussed in ref. \cite{rba}.

\section{The perturbative calculation of the invariant measure
density, $\phi_E (x)$ around the exceptional energy, $E_S$ for the
SRTM} 

The procedure for the calculation is well documented in the
literature. For $E = 2 \cos \alpha \pi$, the calculation is
facilitated by transformation given by equation (\ref{x}). Since, for the
invariant measure density we should have 
\begin{equation}
\phi_E (x) dx = h_E (\theta) d\theta
\end{equation}
We define an operator $J_{\alpha}$ such that 
\begin{equation} \label{j1}
(J_{\alpha} \phi_E ) (\theta) = \phi_E (x) \frac{dx}{d\theta} =
h_E (\theta)
\end{equation}
We also define another operator, $\tau_{\alpha} = J_{\alpha} \; T_E
\; J_{\alpha}^{-1}$ such that 
\begin{equation}\label{j2}
(\tau_{\alpha} \; g) (\theta) = g (\theta - \alpha \pi)
\end{equation}
Furthermore, if $g(x) = \frac{d f(x)}{dx}$, we then have
\begin{equation}\label{j3}
(J_{\alpha} \; G) (\theta) = g(x) \frac{dx}{d\theta} =
\frac{d}{d\theta} \frac{d\theta}{dx} (J_{\alpha} \; \hat f) (\theta)
\end{equation}
where $(J_{\alpha} \; \hat f) (\theta) = f(x) \frac{dx}{d\theta}$. 
It then follows that
\begin{equation}\label{j4}
J_{\alpha} \frac{d}{dx} J_{\alpha}^{-1} = \frac{d}{d\theta}
\frac{d\theta}{dx} = -\frac{d}{d\theta} \;
\frac{\sin^2\theta}{\sin{\alpha \pi}}
\end{equation}

We now consider the RBA and put $E = \omega + \lambda$, where $E = 2
\cos \alpha \pi$ and $\alpha$ is not belong to ${\bf \rm Q}$
\cite{bov}.  It is
transparent from equation (\ref{20}) that the perturbative 
calculation will be less cumbersome if we replace $\omega$ by $(E -
\lambda)$. This yields 
\begin{equation}\label{51}
\phi_{E, \epsilon} (x) = \frac{1}{2} [ T_{E}^3 + T_E \;
T_{E-\epsilon} \; T_E ] \phi_{E,\epsilon} (x)
\end{equation}
when $\epsilon = -\lambda (1 - V_0^2) /V_0^2$. So, the problem
effectively reduces to the calculation of the invariant measure
density of a random binary mixture of two trimeric clusters. The
central energy of the guest cluster is $\epsilon$ and all nearest
neighbor hoppings are unity. We again emphasize that we are seeking 
$\phi_{E, \epsilon} (x) \in L_{+}^{1} ({\bf \dot R}, dx)$ for $\mid E
\mid < 2$. To obtain the perturbative solution of $\phi_{E, \epsilon}
(x)$ around $\epsilon = 0$, we write 
\begin{equation} \label{52}
\phi_{E, \epsilon} (x) = \sum_{n=0}^{\infty} \frac{\epsilon^n}{n!}
\phi_E^{(n)} (x)
\end{equation}
and  $\phi_E^{(n)} (x)$  satisfies $\int_{\bf \dot R}
\phi_{E}^{(n)} (x) dx = \delta_{n,0}$. We further note that
$\phi_E^{0} (x) = \frac{1}{2 \pi} \frac{\sqrt{4-E^2}}{x^2 -E x +1}$.
So, 
\begin{equation}
h_E^{0} (\theta) = \phi_E^{0} (x) \frac{dx}{d\theta} =- \frac{sgn
(\alpha \pi)}{\pi} 
\end{equation}
when $sgn (\alpha \pi)$ is positive or negative depending on whether
$\alpha$ is positive or negative.
We know that according to Anderson's theorem, all eigenstates of the
system will be exponentially localized for $\mid \epsilon \mid > 0$. 
Furthermore, the Lyapunov exponent, $\gamma (E)$ is the inverse
localization length of the eigenstate at $E$. So, physically
$\gamma_{\epsilon} (E) \geq 0$ for $E \in (-2,2)$. 
We also know that $\gamma_{\epsilon} (E)$ is a continuous function of
$\epsilon$ particularly if $\mid \epsilon \mid << 1$. 
So, we can expand $\gamma_{\epsilon_s} (E)$ in a Taylor series around
$\epsilon = 0$ to obtain 
\begin{equation}\label{53}
\gamma_{\epsilon} (E) = \sum_{n=0}^{\infty}
\frac{\epsilon^{n}}{n!} \gamma_{n} (E).
\end{equation}
Since, $\gamma_{0} (E) = 0$, $\gamma (E)$ reaches the minimum value
at $\epsilon = 0$.  
So, for $\mid E \mid < 2$, we must have $\gamma_1 (E) = 0$ and the
leading order term in the Taylor expansion will be $O (\epsilon^2)$
with $\gamma_2 (E) > 0$. We also note in this context that when $\mid
E \mid = 2$, although $\gamma_0 (E) = 0$, however the analytical 
continuation of 
$\gamma (E)$ for $\mid E \mid > 2$ will yield negative $\gamma (E)$.
This naturally follows from the constraint on $\phi_E (x)$. So, in
this case $\gamma_1 (E)$ will be non-zero.  We also see 
from equation(\ref{34}) that $\gamma_{m} (E)$ has no contribution from
$\phi_E^{(m)} (x)$. So, to calculate the leading order term of
$\gamma_{\epsilon} (E)$, the knowledge of $\phi_E^{(1)} (x)$ will
suffice. Similarly for the density of states at $E_S$, we do not need
more than $\phi_E^{(1)} (x)$ [see equation (29)]. 
From equation (\ref{51}) and equation (\ref{52}) for $(J_{\alpha}
\phi_E^{(1)}) (\theta) = h_E^{(1)} (\theta)$ we obtain 
\begin{equation}\label{54}
(\rm I - \tau_{\alpha}^3 ) h_E^{(1)} (\theta) = - \frac{1}{2}
\frac{d}{d\theta} \frac{\sin^2 (\theta - \alpha \pi)}{\sin \alpha
\pi} \tau_{\alpha}^3 \; h_E^{0} (\theta)
\end{equation}
The procedure for solving this type of equations is to expand
$h_E^{n} (\theta)$ in the Fourier series in $(\frac{-\pi}{2},
\frac{\pi}{2})$ and then calculate the coefficients from the governing
equation. This procedure when applied to equation (\ref{54}) yields 
\begin{equation}
h_E^{(1)} (\theta) = \frac{(-1)}{4 \pi} \frac{\cos (2\theta + \alpha
\pi)}{\mid \sin \alpha \pi \mid \sin 3 \alpha \pi}
\end{equation}

The second case that we consider is $V_0 = 1$ and $u = 2 v$. Here,
$v$ and $2 v$ are exceptional energies and we have already shown that
$\gamma (E_S) = 0$ for $E_S = v$ and $2v$ [See equation (33) and equation
(34)]. Since $E_S = E - \lambda$, from equation (\ref{20}) we get for  
$\mid v \mid \leq 1$, 
\begin{equation} \label{55}
\phi_{E(E_S), \lambda} (x) = \frac{1}{2} [ T_E^3 + T_{E - \Sigma_1} \;
T_{E-\Sigma_2} \; T_{E - \Sigma_1}] \phi_{E(E_S), \lambda} (x)
\end{equation}
where $\Sigma_1 = E - \lambda$ and $\Sigma_2 = 2 (E - \lambda)$ if
$E_S = v$. On the other hand, for $E_S = 2 v$, we have $\Sigma_1 =
\frac{E - \lambda}{2}$ and $\Sigma_2 = E - \lambda$. So, here the
effective 
site-energies of the guest cluster depend on the energy under
consideration. When viewed from a broader perspective, such a
situation does arise in the analysis of complex one dimensional
chains like polyaniline, polythiophene etc. But, the situation there
is a bit more complex. In relation to the localization of
eigenstates, arguments presented for the previous example also hold
good here. States which are delocalized for $\lambda = 0$ will be
exponentially localized for $\lambda \neq 0$. Again, the Lyapunov
exponent, $\gamma_{\lambda} (E)$ is expected to be $O (\lambda^2)$.
So, to obtain the first order term in the perturbative calculation of
$\phi_{E(E_S), \lambda} (x)$, we write
\begin{equation}\label{jh1}
(J_{\alpha} \; \phi_{E(E_S), \lambda}) (\theta) = h_{E(E_S), \lambda}
= \sum_{n=0}^{\infty} \frac{\lambda^n}{n!} \; h_{E(E_S)}^{(n)} (\theta)
\end{equation}
Then from equation (\ref{55}) and equation (\ref{jh1}) for $E_S = v$ 
and $2v$ we get 
\begin{equation}\label{jh2}
(\rm I - 2 \tau_{\alpha} + \tau_{\alpha}^4) h_{E(v)}^{(1)} (\theta) =
\frac{E^2}{\pi \mid \sin \alpha \pi \mid} \sin 2\theta
\end{equation}
and 
\begin{equation}\label{ta1}
(2 -  \tau_{\alpha}^3 - \tau_{\alpha})
h_{E(2v)}^{(1)} (\theta) = - \frac{E^2 \sin 2 \theta}{4 \pi \mid \sin
\alpha \pi \mid}
\end{equation}
respectively. Since we plan to calculate only the IDOS for these
cases, we solve for a new function, $f_{E(E_S)} (\theta)$. 
For $E_S = v$ and $2v$ 
\begin{equation}
f_{E(v)} (x) = [{\rm I} - T_E - {T_E}^2 - {T_E}^3] \phi_{E(v)}^{(1)} (x)
\end{equation}
and 
\begin{equation}
f_{E(2v)} (x) = [2 + T_E + {T_E}^2] \phi_{E(2v)}^{(1)} (x)
\end{equation}
Now from the governing equation of $h_{E(E_S)} (\theta)$ [i.e equation
(\ref{jh2}) and equation(\ref{ta1})] we find that  
\begin{equation}\label{sol}
(\rm I -\tau_{\alpha}) \hat f_{E(E_S)} (\theta) = \frac{E^2 \sin
2\theta }{C(E_S) \pi \mid \sin \alpha \pi\mid}
\end{equation}
when $C(S) = 1$ and $-4$ for $E_S = v$ and $2v$ respectively. 
This equation (\ref{sol})can be solved by the standard procedure and 
we obtain 
\begin{eqnarray}\label{gel}
f_{E(E_S)} (x) & = & (J_{\alpha} \hat f_{E(E_S)}) (0) \nonumber \\
& = & \frac{E^3}{4 C(E_S) \sin^2 \alpha \pi} \phi_E^{0}
(x) - \frac{E^3 \pi} {2 C(E_S) \mid \sin \alpha \pi\mid}
{\phi_E^{0}}^2 (x)  \nonumber \\
& & + \frac{E^2}{2 C(E_S)} \frac{d}{dx} \phi_E^{0} (x)
\end{eqnarray}

Finally we note that when $u = (1 + V_0^2) v$ and $V_0^2 (1 - v^2) =
1$, two exceptional energies merge at $E = v$ [$\gamma (v) = 0$ (equation
(33)]. For this case the equation for $\tilde h_{E(v), \lambda}
(\theta)$ [ $\tilde {} $ to note merging of two $E_S$ 's] is 
\begin{equation}
(\rm I - 2 \tau_{\alpha} + \tau_{\alpha}^4) \tilde h_{E(v)}^{(1)}
(\theta) = 0
\end{equation}
The solution for this equation is $\tilde h_{E(v)}^{(1)} (\theta) = {\rm
constant}$. Since 
\begin{equation}\label{imp}
\int_{-\pi/2}^{\pi/2} \tilde h_{E(v), \lambda} (\theta) d\theta = 
\int_{-\pi/2}^{\pi/2} \tilde h_{E(v)}^{(0)}(\theta) d\theta = 1
\end{equation}
by construction, we need 
\begin{equation}
\tilde h_{E(v)}^{(1)}(\theta) = 0
\end{equation}
to satisfy the constraint imposed by equation (\ref{imp}).

\section{Calculation of the Lyapunov exponent}

We consider here the RBA and the SRTM with a degenerate exceptional
energy. We choose the first problem because the guest cluster 
has different hopping elements inside it. On the other hand, 
the second problem allows us to investigate analytically the effect 
of a degenerate exceptional energy on $\gamma (E)$ and the IDOS. 

\noindent (I) The RBA: \\

Since $T_{E-\epsilon} = e^{\epsilon \frac{d}{dx}} \; T_E$, from equation
(32) and equation (55) we obtain
\begin{equation}
\gamma_1 (E) = \frac{1}{6} \int_{-\infty}^{\infty} dx \ln y(x)
\frac{d}{dx} \phi_E^{0} (x) = 0
\end{equation}
and
\begin{eqnarray}
\gamma_2 (E) & = & \frac{1}{6} \int_{-\infty}^{\infty} dx \ln y(x)
(\frac{d}{dx})^2 \phi_E^{0} (x) +
\frac{1}{3} \int_{-\infty}^{\infty} dx \ln y(x)
\frac{d}{dx} T_E^{2} \phi_E^{1} (x) \nonumber \\
& = & \frac{1}{6 \pi \sin^2 \alpha \pi} \int_{-\pi/2}^{\pi/2} 
d\theta \; [ \ln \mid \sin \alpha \pi \mid - \ln \mid \sin \theta \mid ]
(\cos 2\theta - \cos 4\theta) \nonumber \\
&  & - \frac{1}{3 \pi \mid \sin \alpha \pi \mid } \int_{-\pi/2}^{\pi/2} 
d\theta \; [ \ln \mid\sin\alpha \pi \mid - \ln \mid \sin \theta \mid ]
\frac{d}{d\theta} \sin^2 \theta \; \tau_{\alpha}^2 \; h_E^{(1)}
(\theta) \nonumber \\
& = & \frac{1}{12(4-E^2)}
\end{eqnarray}
So, for this case we have 
\begin{equation}
\gamma(\omega + \lambda) \sim \frac{\lambda^2 (1-V_0^2)^2}{24 V_0^4
(4 - \omega^2)} + O (\lambda^3)
\end{equation}
when $\mid \omega \mid < 2$.

\vspace{.25in}

\noindent (II) The SRTM with degenerate exceptional energy : \\

We define an operator, $\tilde {\bf O} (E, \lambda)$ which is the total
operator operating on $\phi_E(v) (x)$ in equation (33). 
\begin{equation}
\tilde {\bf O} (E, \lambda) = T_{\lambda}  T_{-E + \lambda f (E,
\lambda)} T_{\lambda} - T_E - (1-T_E) T_0 [ T_{\lambda} - T_{-E +
\lambda f(E, \lambda)} T_{\lambda}]
\end{equation} 
where $f(E, \lambda) = 2 - E^2 + 2 E \lambda + \lambda^2$. To expand 
$\tilde {\bf O} (E, \lambda)$ around $\lambda = 0$, we write 
\begin{equation}
\tilde {\bf O} (E, \lambda) = \sum_{n=0}^{\infty} \lambda^n {\bf O}_n (E)
\end{equation}
Now using the standard procedure, we obtain 
\begin{eqnarray}
{\bf O}_0 (E) & = &  0 \\
{\bf O}_1 (E) & = & -[\frac{d}{dx} ( T_E + T_E^{-1} + (2-E^2) {\rm I}] \\
{\bf O}_2 (E) & = & \frac{1}{2} (\frac{d}{dx})^2 [ T_E - T_E^{-1} +
(2-E^2)^2 \; {\rm I}] -2 E \frac{d}{dx} + (2-E^2) \frac{d}{dx}
\frac{d}{dx} x^2 \nonumber \\
&  & + \frac{2 \sqrt{1-E^2/4}}{\pi} \frac{d}{dx} \frac{d}{dx}
\frac{1}{\phi_E^{0} (x)} T_E^{-1}
\end{eqnarray}

From equation (33) and equation (66) we find that 
\begin{equation}
\gamma_1 (E) = -\frac{4-E^2}{6} \int_{-\infty}^{\infty} dx \ln y(x)
\frac{d}{dx} \phi_E^{0} (x) = 0
\end{equation}
and 
\begin{eqnarray}
\gamma_2 (E) & = & \frac{1}{6} (2-E^2)^2 \int_{-\infty}^{\infty} dx
\ln y(x) 
(\frac{d}{dx})^2 \phi_E^{0} (x) -
\frac{2}{3} \int_{-\infty}^{\infty} dx \ln y(x)
\frac{d}{dx}  \phi_E^{0} (x) \nonumber \\
&  & + \frac{(2-E^2)}{3} \int_{-\infty}^{\infty} dx \ln y(x)
\frac{d}{dx} \frac{d}{dx} x^2 \phi_E^{0} (x) \nonumber \\
& = & \frac{(2-E^2)^2}{6 \pi \sin^2 \alpha \pi} \int_{-\pi/2}^{\pi/2} 
d\theta [ \ln \mid \sin \alpha \pi \mid - \ln \mid \sin \theta \mid ]
(\cos 2\theta - \cos 4\theta) \nonumber \\
& & + \frac{(2-E^2)}{3 \pi \mid \sin \alpha \pi \mid }
\int_{-\pi/2}^{\pi/2} 
d\theta [ \ln \mid \sin \alpha \pi \mid - \ln \mid \sin \theta \mid ]
[\cos 2(\theta+\alpha \pi) \nonumber \\  
&  & - \cos (4\theta + 2 \alpha \pi)] \nonumber \\ 
& = & 0
\end{eqnarray}
Since $\gamma_2 (v) = 0$ for this case, in principle further
calculation is needed to find the leading non-zero term in the
expansion of $\gamma (E)$. However, the leading term can be obtained
through a simple argument. For any arbitrary $V_0$, the original
system will yield two exceptional energies, $E_{S_1}$ and $E_{S_2}$
provided $\mid E_{S_i} \mid \leq 2$ for $i = 1, 2$. If $\mid v \mid
\leq 1$, so that $V_0^2$ is positive, for $E_{S_2}$ to be the second
exceptional energy we need
\begin{equation}
\frac{1-\mid v \mid}{1 + \mid v \mid} \leq V_0^2 \leq \frac{1 + \mid v
\mid}{1 - \mid v \mid} 
\end{equation}
Now from equation (30) we find that 
\begin{eqnarray}
\gamma (E_{S_2}) & = & \frac{1}{6} \int_{-\infty}^{\infty} dx \ln \mid
x \mid [ T_{E_{S_2} - v} + T_{E_{S_2} - v}^{-1}] \phi_{E_{S_2}}^{0}
(x) \nonumber \\
& = & 0
\end{eqnarray}
Furthermore, $\gamma (v) = 0$.
Since $\gamma (E)$ reaches the minimum value at $v$ and $E_{S_2}$, it
must possess a maximum in between these points. As we bring $E_{S_2}$
towards $v$ by tuning $V_0$, this maximum also moves towards $v$ and
the value of $\gamma (E)$ at the maximum simultaneously reduces. In 
the limit when $V_0^2 (1 - v^2) = 1$, two minima and a maxima merge 
at $v$. So, it is an inflexion point of $\gamma (E)$ and $\frac{d^2 
\gamma}{d E^2} \mid_{E = v}$ should be zero. This is precisely 
obtained. Since for $\mid v \mid < 2$, $\gamma (v)$ must be a 
positive semidefinite quantity with the leading order term 
determining the sign, $\frac{d^3 \gamma}{d E^3} \mid_{E = v}$ must 
be zero and we should have $\gamma (E) \sim (E - v)^4$. This 
prediction can be tested by rigorous 
calculation. We shall, however, present an alternative justification.
Since around the exceptional energies, the system behaves like a weak
disordered system, around these energies $\gamma (E) \sim \mid r (E)
\mid^2$, when $\mid r (E) \mid^2$ is the reflection coefficient of a
simple guest cluster in the host lattice (ref. \cite{rtm}). For this
system $\mid r (E) \mid^2$ can be found in ref. \cite{rtm}. The Taylor
series expansion of $\mid r (E) \mid^2$ around $v$ for $\mid v \mid <
2$ yields 
\begin{equation}
\mid r (E) \mid^2 \sim \frac{v^2 (1-v^2)^2}{(4 - v^2)}
(E-v)^4 + O [(E-v)^5]
\end{equation}
This is consistent with our arguments.

Before concluding this section we show that for the SRTM with $V_0^2
= 1$, $u = 2v$ and $\mid v \mid < 1$, $\gamma_1 (v)$ and $\gamma_2 (2
v)$ are indeed zero. Consider first the case of $v$. From equation (33) 
we obtain 
\begin{eqnarray}
\gamma_1 (E) & = & -\frac{1}{6} \int_{-\infty}^{\infty} dx \ln y(x)
\frac{d}{dx} [T_E^{-1} + T_E + 2] \phi_E^{0} (x) \nonumber \\
& = & -\frac{2}{3} \int_{-\infty}^{\infty} dx \ln y(x) \frac{d}{dx}
\phi_E^{0} (x) \nonumber \\
& = & 0
\end{eqnarray}
On the other hand for $2v$, from equation (34) we obtain 
\begin{eqnarray}
\gamma_1 (E) & = & - \frac{1}{12} \int_{-\infty}^{\infty} dx \ln y(x)
\frac{d}{dx} [ T_E^{-1} + T_E + 2 (\frac{E}{2} - x)^2 T_E ]
\phi_E^{0} (x) \nonumber \\
&  & + \frac{\lambda}{6} \int_{-\infty}^{\infty} dx \ln \mid x \mid
\frac{d}{dx} x^2 T_{E/2} \phi_E^{0} (x) \nonumber \\
& = & -\frac{E^2}{24} \int_{-\infty}^{\infty} dx \ln y(x)
\frac{d}{dx} \phi_E^{0} (x) 
 -  \frac{1}{6} ( 1 - \frac{E^2}{4}) \int_{-\infty}^{\infty} dx \ln
\mid x \mid \frac{d}{dx} T_{E/2} \phi_E^{0} (x) \nonumber \\
& = & 0
\end{eqnarray}
This is so because both integrals involve odd function of $x$. Hence,
around $E_S = v$ and $2v$, $\gamma (E) \sim \frac{\gamma_2}{2} (E -
E_S)^2$. The direct calculation of $\gamma_2$ for these cases albeit
possible, is however quite complicated. But as mentioned earlier a
good estimation of $\gamma_2$ can be obtained from the reflection 
coefficient of the single guest cluster.

\section{Density of states at the exceptional energy}
\noindent (I) The RBA: \\

From equation (29) we get 
\begin{eqnarray}
N (E = \omega + \lambda) & = & 1 - \int_{-\infty}^{0-} dx \phi_E^{0}
(x) - \frac{\varepsilon}{3} \int_{-\infty}^{0-} dx ({\rm I} + T_E + T_E^2)
\phi_E^{(1)} (x) \nonumber \\
&  & - \frac{\varepsilon}{6} \phi_E^{0} (0) + O (\varepsilon^2)
\end{eqnarray}
Again from the governing equation of $h_E^{(1)} (\theta)$, {\it i.e},
equation (54), we obtain 
\begin{eqnarray}
\hat f_E (\theta) & = & (I + \tau_{\alpha} + \tau_{\alpha}^2)
h_E^{(1)} (\theta) \nonumber \\
& = & - \frac{\cos(2 \theta - \alpha \pi)}{4\pi \sin\alpha \pi \mid
\sin \alpha \pi \mid }
\end{eqnarray}
This equation, in turn yields 
\begin{eqnarray}
f(x) & = & \hat f_E (\theta) \frac{d\theta}{dx}
\nonumber \\
& = & \frac{E}{8 \sin^2 \alpha \pi (E^2 - 1)} \phi_E^{0} (x) +
\frac{1}{4(E^2 - 1)} \frac{d}{dx} \phi_E^{0} (x) \nonumber \\ 
&  & - \frac{E \pi}{4 \mid \sin \alpha \pi \mid (E^2 - 1)}
{\Phi_E^{0}}^2 (x) 
\end{eqnarray}
Required integrals for the calculation of $N(E)$ can be easily
performed. The DOS at $\omega$, $\rho (\omega)$ is 
\begin{eqnarray}
\rho (\omega) & = & \frac{d N(E)}{dE} \mid_{E = \omega} \nonumber \\
& = & \frac{1}{\pi \sqrt{4 - \omega^2}} + \frac{(1-V_0^2)}{6 V_0^2 \pi
\sqrt{4 - \omega^2}}  
\end{eqnarray}

\noindent { (II) The SRTM with a degenerate exceptional energy}

We have already shown in this case $\tilde \phi_E^{(1)} (x) = 0$.
Furthermore, to obtain the DOS at $E = v$, we need the coefficient of
$(E - v) = \lambda$ in the Taylor series expansion of $N(E)$. So, we
write from equation (29) 
\begin{eqnarray} \label{85}
N_0 (E = v + \lambda) & = & 1 - \frac{2}{3} \int_{-\infty}^{0-}
\phi_E^{0} (x) dx - \frac{1}{6} \int_{-\infty}^{0-} dx [T_{E-v} +
T_{\frac{E-u}{V_0^2}} \; T_{E-v} ] \phi_E^{0} (x) \nonumber \\
& = & \frac{1}{2} + \frac{2}{3 \pi} \tan^{-1} \frac{E}{\sqrt{4-E^2}} +
\frac{1}{6\pi} \tan^{-1} \frac{E-2v}{\sqrt{4-E^2}} \nonumber \\
&  & + \frac{1}{6\pi} \tan^{-1} \frac{b(E)}{a(E)}
\end{eqnarray}
where 
\begin{displaymath}
b(E) = \frac{E-u}{V_0^2} - \frac{E-2v}{2(v^2 - E v +1)}
\end{displaymath}
and 
\begin{displaymath}
a(E) =  \frac{\sqrt{4-E^2}}{2(v^2 - E v +1)}
\end{displaymath}
Now expanding $N_0 (E)$ in the Taylor series around $E = v$, we get 
for the DOS, $\rho (v)$ 
\begin{equation}
\rho (v) = \frac{1}{\pi \sqrt{4-v^2}} - \frac{v^2}{6 \pi \sqrt{4-v^2}}
\end{equation}

We consider now the case of the SRTM where $V_0^2 = 1$ and $u = 2 v$,
and $\mid v \mid < 1$.
For this case the DOS at $E = v$ and $2v$ has been calculated
numerically. We present here the analytical results. We write for
these cases 
\begin{equation}
N(E= E_S + \lambda) \approx N_0 (E= E_S + \lambda) + \frac{\lambda}{6}
N_1 (E= E_S + \lambda) + O(\lambda^2)
\end{equation}
when $N_0 (E= E_S + \lambda)$ is obtained from equation (\ref{85}). \\
Furthermore, 
\begin{equation}
N_1 (E) = \int_{b(E)}^{\infty} dx f_{E(E_S)} (x)
\end{equation}
and  $b(E) = E$ and 0 for $E_S = v$ and $2v$ respectively. 
$f_{E(E_S)}$ is given by the equation (\ref{gel}). 
After performing the required integral and combining the coefficient 
of $\lambda$ in the expansion of $N(E)$ we get
\begin{eqnarray}
\rho (v) & = & \frac{1}{\pi \sqrt{4-v^2}} + \frac{v^2}
{4 \pi \sqrt{4-v^2}} + \frac{v^2  (1 - 2 v^2)}{24 \pi \sqrt{4-v^2}} 
\end{eqnarray}
and 
\begin{equation}
\rho (2v)  =  \frac{1}{2\pi \sqrt{1-v^2}} - \frac{v^2}{12\pi
\sqrt{1-v^2}} 
\end{equation}

Finally combining the results of the two sections we find that the
number of states having localization length superior to the sample 
size is $\sim \rho (E_S) [\frac{2}{M \gamma_{2n} (E_S)}]^{1/2n}$ where
$M$ is the size of the sample. $n$ denotes the degeneracy of the
exceptional energy. Consequently, the mean square displacement of an
electron should go as $t^{2 \gamma}$ with $\gamma = (1 -
\frac{1}{4n})$. This prediction matches very nicely with exponents
obtained from numerical simulations \cite{rdm,rtm1,pkd}.

\section{One dimensional correlated disordered system as an effective
Lloyd model}

We have already proved that the invariant measure $\phi_E^{0} (x)$ of
the system at the exceptional energies is a Lorentzian distribution 
centered at $E_S /2$ with a half width $\sqrt{1-E_S^2/4}$. We have
also seen through examples that the DOS at $E_S$ in the near perfect
limit to a good approximation is the DOS of the perfect system. It is
also a well established result that the Lyapunov exponent, $\gamma
(E)$ around $E_S$ can be approximated to a fair degree by  $\mid
r(E) \mid^2$ where $\mid 
r(E) \mid^2$ is the reflection coefficient of a single guest cluster 
in the host lattice and $\mid r(E_S) \mid^2 = 0$. 

The Lloyd model \cite{llyod} on the other hand is the uncorrelated 
site disordered Anderson model 
where the probability distribution of the site energies $\{
\epsilon_n \}$, $P (\epsilon_n)$ is 
\begin{equation}
P (\epsilon_n) = \frac{1}{\pi} \frac{\epsilon_1}{(\epsilon_n -
\epsilon_0)^2 + \epsilon_1^2}
\end{equation}
The invariant measure $\phi_E (x)$ for this model is 
\begin{equation}
\phi_E (x) = \frac{1}{\pi} \frac{\epsilon_1^{*}}{(x -
\epsilon_0^{*})^2 + {\epsilon_1^{*}}^2}
\end{equation}
where
\begin{equation}
\epsilon_0^{*} = \frac{E}{2} + \frac{1}{2} {\bf \rm Re} \sqrt{(E + i
\epsilon_1)^2 - 4} = \frac{E}{2} [ 1 + \frac{\tilde
\epsilon_1}{\sqrt{A}}] 
\end{equation}
\begin{equation}
\epsilon_1^{*} = \frac{\sqrt{4-E^2}}{2} + [\tilde \epsilon_1 + \sqrt{A}]
\end{equation}
\begin{equation}
A = \frac{1 + \tilde {\epsilon_1}^2 + \sqrt{(1+\tilde
{\epsilon_1}^2)^2 + 4 E^2 \tilde {\epsilon_1}^2}}{2}
\end{equation}
and 
\begin{equation}
\tilde \epsilon_1 = \frac{\epsilon_1}{\sqrt{4 - E^2}}
\end{equation}
We note that all nearest neighbor hopping matrices have been assumed
to be unity. The Lyapunov exponent and the DOS for this model can be
found in the literature \cite{mat}. We simply quote the results.
\begin{equation}\label{ll0}
4 \cosh \gamma (e) = \mid 2 + E \mid [ 1 + \frac{2-E}{2+E} \tilde
\epsilon_1^2]^{1/2} + \mid 2 - E \mid [ 1 + \frac{2+E}{2-E} \tilde
\epsilon_1^2]^{1/2} 
\end{equation}
and 
\begin{equation}
\rho (E) = \frac{1}{\pi} e^{2 \gamma (E)} [ \epsilon_1^{*}
\frac{d\epsilon_0^{*}}{dE}  - \epsilon_0^{*} \frac{d\epsilon_1^{*}}{dE}]
\end{equation}
Now in the limit $\tilde \epsilon_1 \rightarrow 0$, we find that 
$\epsilon_0^{*} (E) = \frac{E}{2}$, $\frac{d\epsilon_0^{*} (E)}{dE} =
\frac{1}{2}$, $\epsilon_1^{*} (E) = \frac{1}{2} \sqrt{4 - E^2}$ and 
$\frac{d\epsilon_1^{*} (E)}{dE} = - \frac{E}{2 \sqrt{4 - E^2}}$.
Consequently, we obtain 
\begin{equation}
\phi_E (x) = \frac{1}{\pi} \frac{\sqrt{1-E^2/4}}{x^2 - Ex +1}, 
\end{equation}
\begin{equation}
\gamma (E) = 0
\end{equation}
and 
\begin{equation}
\rho (E) = \frac{1}{\pi \sqrt{4 - E^2}}
\end{equation}
So, all the characteristic features of one dimensional correlated
disordered systems around the exceptional energies are recovered. To
develop an effective Lloyd model for one dimensional correlated
disordered systems, we need then $\tilde \epsilon_1 = f (E)$ such
that $f (E_S) = 0$ for $\mid E_S \mid \leq 2$ and $f(E)$ should also 
contain in it the information about the guest cluster. We further
note that for $\tilde \epsilon_1 \sim 0$, equation (\ref{ll0}) yields
$\gamma (E) \sim \mid \tilde \epsilon_1 \mid$. But we have shown that
for a nondegenerate exceptional energy, $\gamma (E) \sim (E -
E_S)^2$. So, $f (E)$ should also satisfy this condition. In as much
as $\mid r(E)\mid^2$ satisfies all these criteria, we propose that
$\tilde \epsilon_1 = \mid r(E) \mid^2$. This proposal in turn implies
that 
\begin{equation}
P (\epsilon) = \frac{1}{\pi} \frac{\sqrt{4-E^2} \mid r(E)
\mid^2}{\epsilon^2 + (4-E^2) \mid r(E)\mid^4}
\end{equation}
in the effective Lloyd model. Finally we note that many methods
\cite{poly} are
developed to study the electrical conductivity of uncorrelated site 
disordered system. So, these methods can be applied to the systems
considered here through the proposed mapping. 

\section{Summary}
The behavior of electronic states of one dimensional correlated
disordered systems around exceptional energies ($E_S$) is studied
analytically using the invariant measure method. The RDM is the
simplest example in this category and this has been studied by 
this method by Bovier. The basic approach of Bovier is generalized 
thoroughly and rigorously to take into consideration more structure
in the guest cluster. The formalism is further elaborated by applying
to the SRTM. 
Another useful contribution is the alternative mathematical definition 
of the exceptional energy from the invariant measure. This definition
is further substantiated by physical arguments. Furthermore, from
our definition of exceptional energy we obtain an equation
constraining the parameters of the guest cluster. The same equation
has also been obtained by setting $\mid r(E)\mid^2 = 0$. This clearly
shows an intimate relationship between the invariant measure and
$\mid r(E) \mid^2$. This relationship is further highlighted here by
mapping these systems to an effective Lloyd model. This equation
further shows how the structure of the guest cluster can be modulated
to tune positions of exceptional energies. We also obtain through
it the condition for a degenerate exceptional energy. Hence, the
importance of the method is further illustrated. 

In relation to the localization of eigenstate we find expectedly
$\gamma (E) \sim (E - E_S)^2$ for cases with a non degenerate
exceptional energy. In the case of a degenerate resonance, $\gamma_2
(E_S)$ along with $\gamma_1 (E_S)$ are shown to be identically zero.
These results are further substantiated by rigorous analytical
arguments. Further analytical arguments are presented to show that
$\gamma (E) \sim (E - E_S)^4$ for this case. Although a system
containing a degenerate exceptional energy is studied previously by
us, this is, however, the most rigorous analysis. 

In the RDM $\rho (v)$ is the DOS of the perfect system at $E = v$.
Further structure in the guest cluster is found to be manifested 
in $\rho (E_S)$ through correction terms. However, the universality 
in the mathematical expression of $N_1(E)$ should not be overlooked.
This exemplifies further the universality of one dimensional
correlated disordered systems around exceptional energies. We finally
add that the real significance of our work along with that of Bovier 
on the RDM 
is that this firmly establish the anomalous behavior of one
dimensional correlated disordered systems around exceptional energies.

\end{document}